\documentclass[]{raa}            
\usepackage{graphicx,times}
\usepackage{natbib}

\providecommand{\bm}[1]{\mbox{\boldmath$#1$\unboldmath}}

\begin{document}

   \title{Structure of ADAFs in a general large-Scale
B-field:\\
The role of wind and thermal conduction
}

   \volnopage{Vol.0 (200x) No.0, 000--000}      
   \setcounter{page}{1}          

   \author{A. Mosallanezhad
      \inst{1,2}
      \and M. Khajavi
      \inst{3}
   \and S. Abbassi
      \inst{2,4}
   }

   \institute{Center for Excellence in Astronomy \& Astrophysics (CEAA - RIAAM) - Maragha, IRAN, P. O. Box: 55134 - 441,\\
        \and
             School of Physics, Damghan University, P.O.Box 36715-364, Damghan, Iran\\
        \and
        Department of Physics, School of Sciences, Ferdowsi University of Mashhad, Mashhad, 91775-1436, Iran\\
        \and
             School of Astronomy, Institute for Research in Fundamental Sciences, P.O.Box 19395-5531, Tehran, Iran; {\it abbassi@ipm.ir}\\
   }

   \date{Received~~2009 month day; accepted~~2009~~month day}

\abstract{ We have examined the structure of hot accretion flow with a large-scale
magnetic field. The importance of outflow/wind and thermal conduction
on the self-similar structure of a hot accretion flows has been investigated.
In comparison to the
accretion disk without winds/outflow, our results show that the radial and rotational
velocities of the disk become faster however it become cooler because of the
angular momentum and energy flux which are taking away by the winds/outflows. but
thermal conduction opposes the effect of winds/outflows not only decrease the rotational
velocity but also increase the radial velocity as well as the sound speed of
the disk. In addition we have studied the effect of global magnetic field on the
structure of the disk. We have found out that all three components of
magnetic field have a noticeable effect on the structure of the disk such as velocities and
vertical thickness of the disk.
}

   \authorrunning{A. Mosallanezhad, M. Khajavi, S. Abbassi}            
   \titlerunning{Structure of ADAFs in a general large-Scale B-field}  

   \maketitle

%
%
\section{Introduction}           
\label{sect:intro}
Black hole accretion disks provide the most powerful energy-production mechanism in the universe.
It is well accepted that many astrophysical systems are powered by black hole accretion.
Most of the observational features of black hole accretion systems can be explained through The standard thin disks ( geometrically thin and optically thick accretion disks) with a significant success
(Shakura \& Syunyaev, 1973). The ultraviolet and optical
emission which is observed in quasars frequently has a role in the thermal
radiation in the standard disks (SD) that surrounds by the massive black holes
in quasars (e.g., Sun \& Malkan, 1989). On the other hands, the SD seems unable
to recreate the spectral energy distributions (SED) of in a lot of other luminance sources (e.g.,
Sgr A.) accreting in low rates, and the advection dominated accretion flows
(ADAF) were recommended to be present in these sources (Narayan \& Yi, 1994,
1995). In the ADAF model, the most energy which is released from the infalling gases in
the accretion flow is changed to the internal energy of the gas. only a small fraction of the
energy in ADAFs gets radiated away, thus, their radiation efficiency is
a lot less than that for SD (see Narayan et al., 1998,
for a review and references therein,  Kato, Fukue \& Mineshige 2008).

Outflows/wind have been found in numerical simulations of
hot accretion flow by both magnetohydrodynamic (MHD) and hydrodynamic (HD)
simulations. In some pioneer paper outflow was reportedin HD simulations are by
Igumenshchev \& Abramowicz (1999, 2000) and Stone, Pringle \& Begelman
(1999). Igumenshchev \& Abramowicz (2000) have shown that convective accretion flows and flows with large-scale circulations have significant outward-directed energy fluxes, which have important implications for the spectra and luminosities of accreting black holes. Most recent work focusing on the convective outflows is by Yuan \& Bu (2010). As a result, they have shown that the mass accretion rate decreases inward, i.e., only a small fraction of accretion gas
can fall onto the black hole, while the rest circulates in the convective eddies or lost in convective outflows. Stone \& Pringle (2001) in their MHD simulations  have reported the appearance of wind and outflow. In their simulations the net
mass accretion rate is small compared to the mass inflow and outflow rates at large
radii associated with turbulent eddies.

There are a lots of models indicate that modeling the
hot accretion flows is a big challenging and it is controversial problem which should be investigated.
Thermal conduction is one of the important neglected physical
phenomena in the modeling of ADAFs. Recent observations
of the hot accretion flow around AGNs
indicated that they should be based on collision-less regime (Tanaka \& Menou 2006).
Chandra observations provide tight constraints on the physical parameters of gas.  Tanaka
\& Menou (2006) have used these constraints ( Loewenstein et al. 2001;
Baganoff et al. 2003; Di Matteo et al. 2003; Ho et al.
2003) to calculate the mean free path for the gas materials.
They have reported and suggested that thermal conduction can be as a possible
choice as a mechanism for transporting sufficiently extra heating is provided
in hot advection dominated accretion flows.
So it should be important to consider the role of thermal
conduction in a ADAF solution. Shadmehri (2008),  Abbassi et al. 2008,2010,
Tanaka \& Menou (2006), Faghei 2012, have studied the effect of hot accretion flow with thermal conduction with a semi-analytical method;
dynamics of such systems have been studied in simulation models (e. g. Sharma et al. 2008; Wu et al.
2010). Shadmehri (2008) has shown the thermal conduction oppose the
rotational velocity, but increase the temperature. Abbassi et al. (2008) have shown that for this problems there are two types of
solution, high and low accretion rate. They had plotted the radial velocity for both of the solutions which have shown that
it will modified by thermal conduction.

previous studies about black hole accretion disks recommend that a large-scale magnetic field root in the ISM or even central engine, will be taken inward and compressed closed in the black hole by the accreting
materials (Bisnovatyi-Kogan \& Ruzmaikin 1974, 1976). Therefor, large-scale magnetic field has a great role on the structure of hot flow because of 
highly ionized flow. The effect of magnetic field on the structure of ADAFs were also reported by ( Balbus \& hawley 1998, Kaburiki 2000, Shadmehri 2004, Meier 2005, Shadmehri \& khajenabi 2005, 2006, Ghanbari et al. 2007, Abbassi et al. 2008, 2009, Bu,Yuan \& Xie (2009)). Akizuki \& fukue 2006 and Abbassi et al 2008, presented the self-similar solutions of the flow based on the vertically integrated equations. At the same time, they have stressed on the intermediate case where the magnetic force is compared with other forces by this assumption that the physical variables of an accretion disk have just radius function and also they explained more about disk toroidal magnetic field .

In this paper, we first extend the work of Akizuki \& Fukue 2006; Zhang \& Dai 2008 and Abbassi et al. 2008, 2010 by considering a general large-scale magnetic field in all the three components in cylindrical coordinates $ (r, \varphi, z) $ and then discuss
effects of the global magnetic field on the flows with thermal conduction and wind. We adopt the treatment
that the flow variables are functions of the disk radius, neglect the different structure in
the vertical direction except for the z-component momentum equation. We also compare our results with pervious studies, in which a large-scale magnetic field, thermal conduction and wind are neglected.

\section{Basic Equations}
\label{sect:Obs}

We are interested in analyzing the structure of a magnetized ADAF with a large-scale magnetic field where thermal conduction and wind play an important role in energy and angular momentum transportation. Then, we consider a accreting and rotating disk around a compact
black hole of mass $ M_* $. So, for a steady axi-symmetric
accretion flow,(i.e., $ \partial/\partial t = \partial/\partial \varphi=0  $),
we write our equations in the cylindrical
coordinates $ (r, \varphi, z) $. We vertically integrated
the equations and, then all our physical variables become
only function of radial distance $ r $. Moreover the disk is supposed to
have Newtonian gravity in radial direction and also we neglect
the general relativistic effects. we suppose that the disc have magnetic
field with three components $ B_r $, $ B_{\varphi} $ and $ B_z $.

The continuity equation reads,
\begin{equation}\label{continuity1}
    \frac{\partial}{\partial r}(r \Sigma v_r)+ \frac{1}{2 \pi}\frac{\partial \dot{M}_w}{\partial r} = 0
\end{equation}
where $ v_r $ is radial velocity and $ \Sigma $ is
the surface density at the cylindrical radius $ r $,
which we can define as $ \Sigma = 2 \rho H $,  $ H $ is the half-thickness and $ \rho $ being
midplane density of the disk. The mass loss rate by wind/outflow
is represented by $ \dot{M}_w $, so
\begin{equation}\label{Mdot1}
    \dot{M}_w(r) = \int 4 \pi r^{\prime}\dot{m}_w(r^{\prime})d r^{\prime}
\end{equation}
Here $ \dot{m}_w(r) $ is the mass-loss rate per unit area from each
disk face. According to Blandford \& Begelman 1999; Shadmehri 2008,
We write the dependence of accretion rate as follows

\begin{equation}\label{Mdot2}
    \dot{M} = -2 \pi r \Sigma v_{r} = \dot{M}_{0}(\frac{r}{r_{0}})^{s}
\end{equation}
where $ \dot{M}_{0} $ is the mass accretion rate of the disk material at the outer edge of the disk $ r_{0} $ and $ s $ with s are constant of order unity (Blandford \& Begelman 1999). For a disk with outflow/wind, $ s > 0 $ while in the absence of wind/outflow, we consider $ s = 0 $ (Fukue 2004).

By considering equations (\ref{continuity1})-(\ref{Mdot2}), we
can have

\begin{equation}\label{mdot_wind}
    \dot{m}_{w} = s \frac{\dot{M}_{0}}{4 \pi r^{2}_{0}} (\frac{r}{r_{0}})^{s - 2}
\end{equation}

For the equation of motion in the radial direction we can write,
\begin{equation}\label{motion_r}
     v_r \frac{dv_r}{dr} = \frac{v^2_{\varphi}}{r} - \frac{GM_*}{r^2} - \frac{1}{\Sigma}\frac{d}{dr}(\Sigma c^2_{s}) - \frac{1}{2\Sigma}\frac{d}{dr}(\Sigma c^2_{\varphi} + \Sigma c^2_{z})\\
      - \frac{c^2_{\varphi}}{r}
\end{equation}
Here, $ c_s $ the
isothermal sound speed, which it can define as $ c^2_{s} \equiv p_{gas}/\rho $, $ p_{gas} $
being the gas pressure and $ v_{\varphi} $ is the rotational velocity. Also, $ c_r $, $ c_{\varphi} $ and
$ c_z $ are Alfv$\acute{e}$n sound speeds in three direction of cylindrical
coordinate and define as, $ c^2_{r, \varphi, z} \equiv B^2_{r, \varphi, z}/(4 \pi \rho) $.

We suppose that, just  $\varphi$-component of viscose stress
tensor is important which is define as $ t_{r\varphi}=\mu r d\Omega/dr $ ,
where $ \mu (\equiv \nu \rho) $ is the viscosity and $ \nu $ is the kinematic
constant of viscosity. So we have
\begin{equation}\label{nu}
    \nu = \alpha c_{s} H
\end{equation}
Here $ \alpha  $ is the standard viscous parameter which is  less than unity and constant
(Shakura \& Sunyaev 1973).

 The angular transfer equation with considering outflow/wind can be written as
\begin{equation}\label{motion_phi}
    \frac{v_r}{r}\frac{d}{dr}(r v_{\varphi}) = \frac{1}{r^2 \Sigma}\frac{d}{dr}(\nu \Sigma r^3 \frac{d\Omega}{dr}) + \frac{c_r}{\sqrt{\Sigma}}\frac{d}{dr}(\sqrt{\Sigma} c_{\varphi})\\
    + \frac{c_r c_{\varphi}}{r}- \frac{l^2 \Omega}{2 \pi \Sigma}\frac{d \dot{M}_w}{dr}
\end{equation}
The last term on the right side of above equation represents angular momentum
carried by the wind/outflow. As we know, $ l = 0 $ corresponds to
a non rotating outflow/wind and $ l=1 $ to outflowing material that carries
 the specific angular momentum away(e.g., Knigge 1999).

Similarity, by integrating over $ z $ of hydrostatic balance , we will have
\begin{equation}\label{motion_z}
    \Omega^2_{K} H^2 - \frac{1}{\sqrt{\Sigma}}c_{r} \frac{d}{dr}(\sqrt{\Sigma} c_{z})H = c^2_{s} + \frac{1}{2}(c^2_{r} + c^2_{\varphi})
\end{equation}

Now we write the energy equation considering heating and cooling
processes in an hot accretion flow. We suppose that the generation energy due to viscous
dissipation and  are balanced by the
advection cooling and energy loss of wind/outflow. Thus
\begin{equation}\label{energy}
    \frac{\Sigma v_{r}}{\gamma - 1}\frac{d c^2_{s}}{dr} -2 H v_{r} c^2_{s} \frac{d \rho}{dr} = f \nu \Sigma r^{2}(\frac{d \Omega}{dr})^{2} - \frac{2H}{r}\frac{d}{dr}(r F_{s})\\
     - \frac{1}{2}\eta \dot{m}_{w}(r)v^2_{K}(r)
\end{equation}
where the second term on right hand side of the energy  equation represents transfer of energy
due to the thermal conduction and according to Cowie \& Makee 1977, $  F_{s} ( = 5 \phi_{s} \rho c^3_{s}) $
will be the saturated conduction flux.  The
coefficient $ \phi_{s} $ is less than unity. Here, the last term on
right hand side of energy equation shows the energy loss due to
outflow/wind.
In our case, we consider the dimensionless parameter $ \eta $
it a free parameter and constant (Knigge 1999).

To measure the magnetic field escaping/creating rate,
we can write the three components of induction equation,
$ (\dot{B}_{r}, \dot{B}_{\varphi}, \dot{B}_{z} ) $, as
\begin{equation}\label{induction1}
    \dot{B}_{r} = 0,
\end{equation}
\begin{equation}\label{induction2}
    \dot{B}_{\varphi} = \frac{d}{dr}(v_{\varphi} B_{r} - v_{r} B_{\varphi}),
\end{equation}
\begin{equation}\label{induction3}
    \dot{B}_{z} = -\frac{d}{dr}(v_{r} B_{z}) - \frac{v_{r} B_{z}}{r}.
\end{equation}

Here $\dot{B}_{r,\varphi, z}$ is the field creating or scaping rate due to dynamo effect or magnetic instability.
Now we have a set of magnetohydrodynamical equations (MHD), that describe the structure of magnetized advection dominated accretion flows (ADAFs). The solutions of these equations are related to viscosity, magnetic field strength $\beta_{r,\varphi,z}$, the advection parameter $f$, the thermal conduction parameter $\phi_s$ and wind/outflow parameter $s$ of the disks. In the next section, we will seek a self-similar solution for our main equations.

\section[]{Self-Similar Solutions}

To better understanding of the physical processes taking place in our hot accretion disk, we seek self-similar solutions of the above equations. Therefore, following Narayan \& Yi (1994), we write
similarity solutions as

\begin{equation}\label{self_sigma}
    \Sigma(r) = c_{0} \Sigma_{0} (\frac{r}{r_{0}})^{s-\frac{1}{2}}
\end{equation}

\begin{equation}\label{self_vr}
    v_{r}(r) = - c_{1} \sqrt{\frac{GM_*}{r_{0}}}(\frac{r}{r_{0}})^{-\frac{1}{2}}
\end{equation}

\begin{equation}\label{self_phi}
    v_{\varphi}(r) =  c_{2} \sqrt{\frac{GM_*}{r_{0}}}(\frac{r}{r_{0}})^{-\frac{1}{2}}
\end{equation}

\begin{equation}\label{self_cs}
    c^{2}_{s}(r) =  c_{3} (\frac{GM_*}{r_{0}})(\frac{r}{r_{0}})^{-1}
\end{equation}

\begin{equation}\label{self_cmag}
    c^{2}_{r,\varphi,z}(r) =  \frac{B^{2}_{r,\varphi,z}}{4 \pi \rho} = 2 \beta_{r,\varphi,z} c_{3} (\frac{GM_*}{r_{0}})(\frac{r}{r_{0}})^{-1}
\end{equation}

\begin{equation}\label{self_H}
    H(r) = c_{4} r_{0}(\frac{r}{r_{0}})
\end{equation}
Here ,$ c_{0} $, $ c_{1} $, $ c_{2} $, $ c_{3} $ and $ c_{4} $ are constants and will be determined
later. $ r_{0} $ and $ \Sigma_{0} $ are made the equations in non-dimensional forms. Moreover
the constants
$ \beta_{r,\varphi,z} $ measure the radio of
the magnetic pressure in three direction to the pressure of gas (i.e.,
$ \beta_{r,\varphi,z} = p_{mag,r,\varphi,z}/p_{gas} $).

The field creating/scaping rate $\dot{B}_{r,\varphi, z}$ is assume to be as

\begin{equation}\label{self_Bdot}
    \dot{B}_{r,\varphi,z} = \dot{B}_{r0,\varphi0,z0} (\frac{r}{r_{0}})^{\frac{1}{2}(s - \frac{11}{2})}
\end{equation}
where $ \dot{B}_{r0,\varphi0,z0} $ are constant.

If we substitute the above self-similar solutions in to the main disk equation such as,
the continuity, momentum, angular momentum, hydrostatic
balance and energy equation of the disk, we can obtain the following
system of equations to be solved for $ c_{0} $, $ c_{1} $,$ c_{2} $
$ c_{3} $ and $c_{4}$:

\begin{equation}\label{mdot}
    c_{0}c_{1} = \dot{m}
\end{equation}

\begin{equation}
	-\frac{1}{2} c^2_{1} = c^2_{2}  - 1  - \big[(s - \frac{3}{2})+(s - \frac{3}{2})\beta_z + (s + \frac{1}{2})\beta_{\varphi} \big]c_3
\end{equation}
\begin{equation}\label{2}
    - \frac{1}{2} c_1 c_2 = -\frac{3}{2} (s + \frac{1}{2}) \alpha c_2 \sqrt{c_3} c_{4} +(s +  \frac{1}{2}) c_3 \sqrt{\beta_r \beta_{\varphi}}\\
  - sl^{2} \frac{\dot{m}}{c_{0}}c_{2}
\end{equation}
\begin{equation}\label{3}
    c_{4} = \frac{1}{2}\bigg[\sqrt{(s - \frac{3}{2})^2 \beta_r \beta_{z} c^2_3 + 4 (1 + \beta_r + \beta_{\varphi}) c_3}\\
    + (s - \frac{3}{2})\sqrt{\beta_r \beta_{z}}c_3 \bigg]
\end{equation}
\begin{equation}\label{motinvarphi}
\big( \frac{1}{\gamma - 1} +s - \frac{3}{2}\big) c_1 c_3 = \frac{9}{4}\alpha f c^2_{2} \sqrt{c_3} c_{4} - 5 \phi_s (s - 2) c^{\frac{3}{2}}_{3} \\
 - \frac{s}{4} \eta \frac{\dot{m}}{c_{0}}
\end{equation}

where $ \dot{m} $ is the non dimensional mass accretion rate which can write as

\begin{equation}
	\dot{m} = \frac{\dot{M}_{0}}{2 \pi r_{0} \Sigma_{0} \sqrt{GM_*/r_{0}}}
\end{equation}
 Also, for the magnetic field creating/scaping rate, we have,
\begin{equation}\label{inducton11}
    \dot{B}_{0r} = 0
\end{equation}
\begin{equation}
\dot{B}_{0\varphi} = \frac{1}{2}(s -\frac{7}{2}) \frac{GM_*}{r^{5/2}_{0}} \sqrt{\frac{4 \pi c_{0} c_{3} \Sigma_0}{ c_{4} }} \times\\
\big(c_2 \sqrt{\beta_r}+ c_1 \sqrt{\beta_{\varphi}} \big)
\end{equation}
\begin{equation}\label{inducton33}
     \dot{B}_{0z}  = \frac{1}{2}(s - \frac{3}{2})  c_1 \frac{GM_*}{r^{5/2}_{0}}   \sqrt{\frac{4 \pi  \beta_{z} c_{0} c_{3} \Sigma_0}{ c_{4} }}
\end{equation}
Without mass outflow/wind, thermal conduction and magnetic
field,(i.e. $ s =  l = \eta = \phi_{s} = \beta_{r} = \beta_{\varphi} = \beta_{z} = 0 $),
our equations and their solutions are reduced to the standard ADAFs
solution (Narayan \& Yi 1994). Moreover in the absence of outflow/wind and thermal conduction
they are reduced to Zhang \& Dai 2008.

Now we can analysis behavior of the solutions in the presence of large-scale magnetic field, outflow/wind and
thermal conduction. The parameters of this
model are the advection
parameter $ f $, the radio of the specific heats $ \gamma $,
the standard viscose parameter $ \alpha $, the mass loss
parameter $ s $, the degree of magnetic pressure to the pressure of gas
, $ \beta_{r} $,
$ \beta_{\varphi} $ and $ \beta_{z} $, the  conduction parameter
$ \phi_{s} $ and $ l, \eta $ parameters corresponding to the non-rotation and rotation
 wind/outflow and the loss of energy by wind/outflow.
\begin{figure*}[tb]
\centering
\includegraphics[width=15cm]{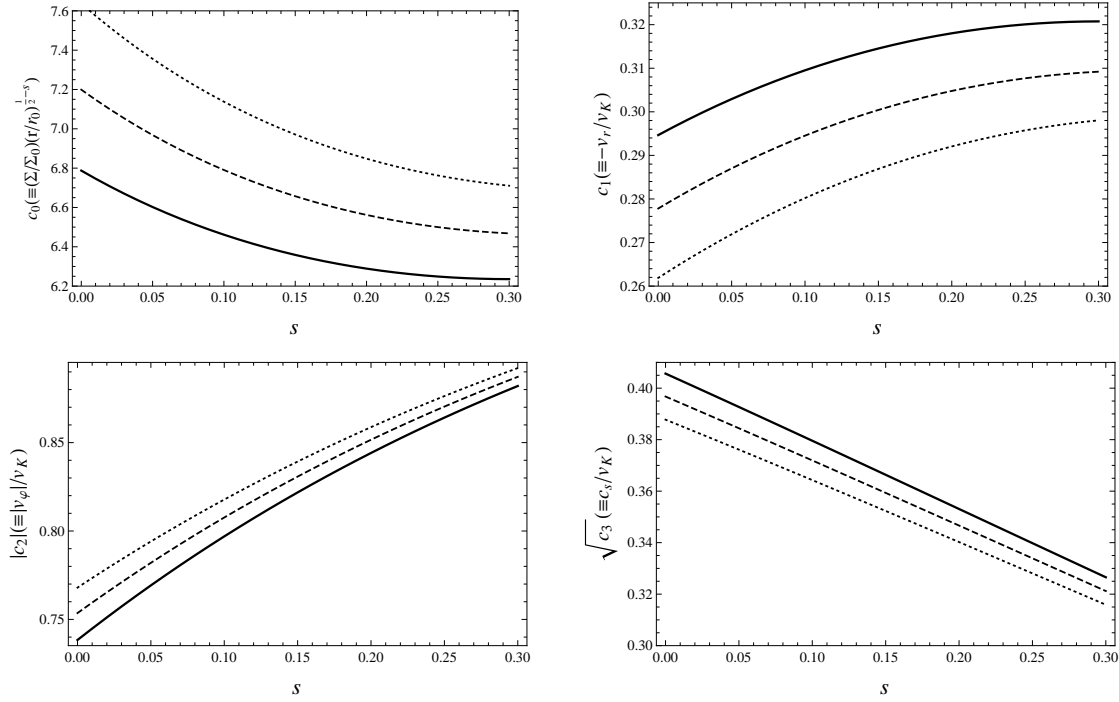}
\caption{Numerical coefficient $ c_{i} $s as a function of
wind parameter $ s $ for several values of thermal conduction parameter $ \phi_{s} $. The dotted, dashed and solid
lines correspond to $ \phi_{s} = 0.001, 0.01 $ and $ 0.02 $ respectively. Parameters
are set as  $ \alpha = 0.2 $, $ \gamma=4/3 $, $ \beta_{r}=\beta_{\varphi} = \beta_{z} = 1.0 $,
$ \eta = l = 0.1 $ and $ f = 1 $.}
\label{thermal}
\end{figure*}

 Four panels in figure \ref{thermal} show the variations of coefficients
$ c_{0} $, $ c_{1} $, $ c_{2} $ and $ c_{3} $ in term of wind
parameter $ s $ for different values of thermal conduction parameter $ \phi_{s} $, i. e.,
$ \phi_{s} = 0.001 $ (dotted line), $ \phi_{s} = 0.01 $ (dashed line) and $ \phi_{s} = 0.02 $ (solid line),
corresponding to $ \alpha = 0.2 $, $ \gamma = 4/3 $, $\beta_{r}= \beta_{\varphi} = \beta_{z} = 1.0 $,
$ \eta = l = 0.1 $  and $ f = 1.0 $ (fully advection). In upper-left
panel of figure \ref{thermal} we can see that for non-zero $ s $, the
surface density is lower than the standard ADAF solution and for stronger
outflows, the reduction of the surface density is more evident . Also surface density decreases by
 increasing of thermal conduction. Upper-right panel of figure \ref{thermal}
represents the behavior of radial infall velocity to
the Keplerian one. Although radial velocity is slower than
the Keplerian velocity, it becomes faster by increasing  $ f $ and $ \alpha $.
We know that the value of $ s $ measure the strength of outflow/wind, and large
values of $ s $ denotes strong wind/outflow. So we can see that, the redial flows
of the accretion materials become faster by increasing of wind/outflow parameter $ s $.
Moreover thermal conduction increases the accretion velocity.
\begin{figure*}[tb]
\centering
\includegraphics[width=15cm]{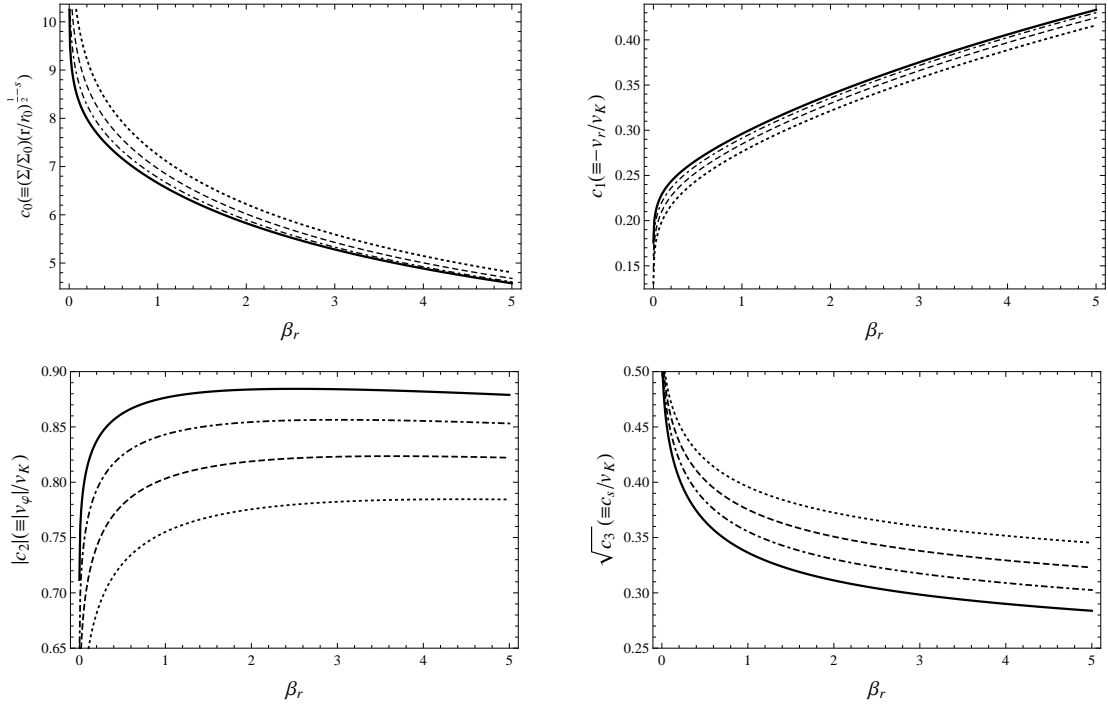}
\caption{Numerical coefficient $ c_{i} $s as a function of magnetic parameter
$ \beta_{r} $ for several values of wind parameter $ s $. The dotted, dashed, dash-dotted and solid
lines correspond to $ s = 0.0, 0.1, 0.2 $ and $ 0.3 $ respectively. Parameters
are set as $ \alpha = 0.2 $, $ \gamma=4/3 $, $ \beta_{\varphi} = \beta_{z} = 1.0 $,
$ \phi_{s} = 0.01 $, $ \eta = l = 0.1 $ and $ f = 1 $.}
\label{beta-r}
\end{figure*}
It is clear from  bottom panels of Figure \ref{thermal}, the rotational velocity
increase by adding wind parameter, $ s $. Although thermal conduction decreases
the rotational speed. It is shown that temperature decreases for strong outflows (down-right panel).
On the other hand, outflow/wind plays as a cooling agent (Shadmehri 2008; Abbassi et al 2010).

Figure \ref{beta-r} shows how the coefficients $ c_{i}s $ depend on
the magnetic parameter in radial direction $ \beta_{r} $ for different
values of outflow/wind parameter
$ s $. We can see that the surface density become larger if the wind parameter $ s (0-0.3) $ increases.
It is obvious that $ s = 0 $ represent no-wind solutions (Tanaka \& Menou 2006). By adding $ \beta_{r} $,
we see that the surface density and sound speed decrease. While, the radial and rotational velocity increase.
The same as figure \ref{thermal}, outflow increase radial and rotational velocity, although the sound speed and
surface density decreases by increasing of wind parameter.

\begin{figure*}[tb]
\centering
\includegraphics[width=15cm]{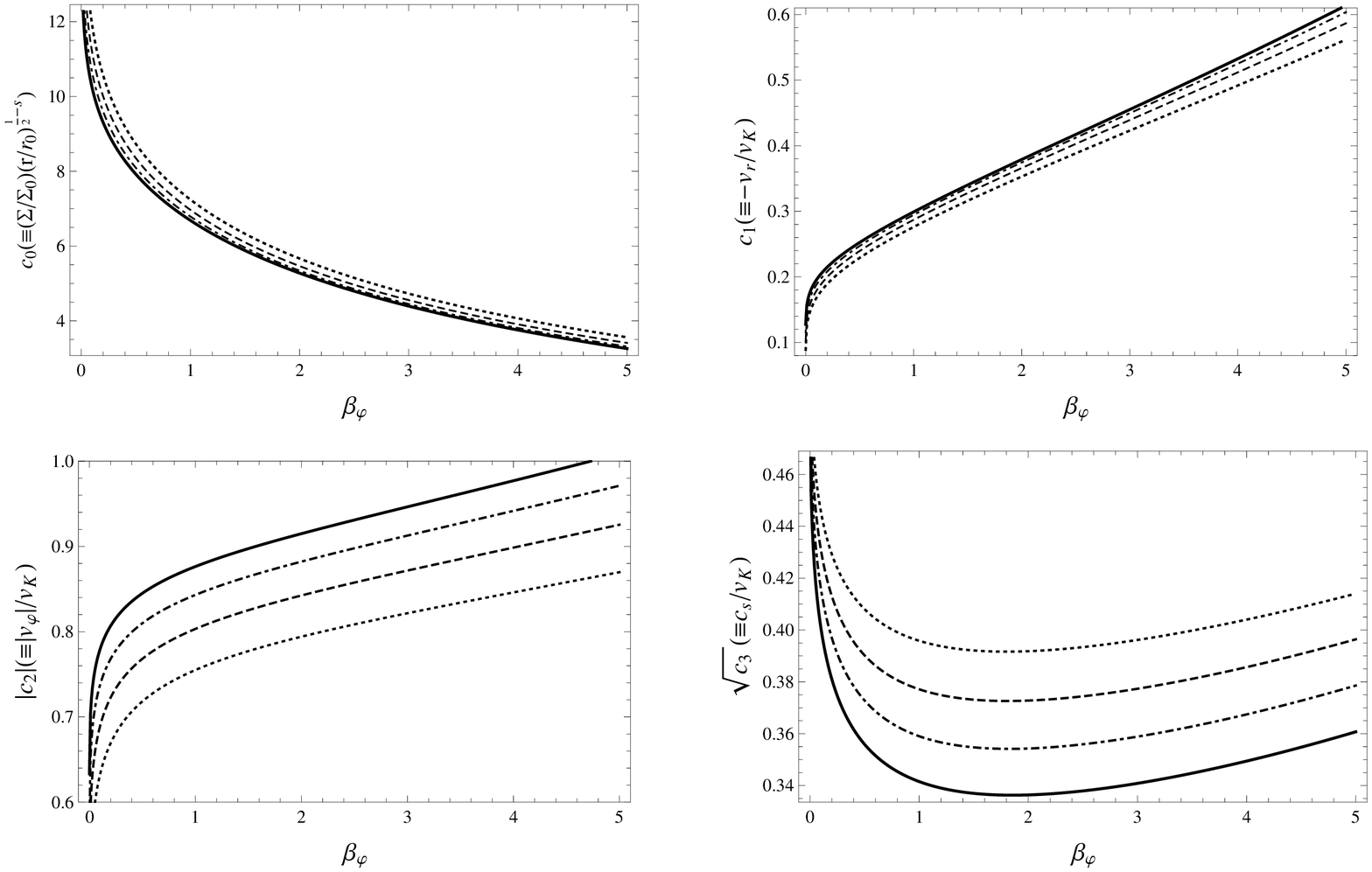}
\caption{Numerical coefficient $ c_{i} $s as a function of magnetic parameter
$ \beta_{\varphi} $ for several values of $ s $. The dotted, dashed, dash-dotted and solid
lines correspond to $ s = 0.0, 0.1, 0.2 $ and $ 0.3$ respectively. Parameters
are set as $ \alpha = 0.2 $, $ \gamma=4/3 $, $ \beta_{r} = \beta_{z} = 1.0 $,
$ \phi_{s} = 0.01 $, $ \eta =l = 0.1 $ and $ f = 1 $.}
\label{beta-phi}
\end{figure*}

In figure \ref{beta-phi} the behavior of the coefficients $ c_{i}s $ versus toroidal magnetic field are plotted
for different values of wind parameter $ s $. From the upper-left panels of figure \ref{beta-phi},
we can see that for $ \beta_{\varphi} (=0-5)$, the accretion velocity is sub-Keplerian, and
that becomes faster by increasing of wind/outflow parameter $ s $. Moreover, when the
toroidal magnetic field becomes stronger, the accretion velocity of flow increases.
This is because the magnetic tension terms, which dominates the
magnetic  pressure term in the radial momentum equation that assists
the motion of accreting materials in radial direction.
The down-right panel in figure \ref{beta-phi} display the rotational velocity of accretion disc.
We can see that for the given advection parameter $ f(=1.0) $  and also $ \beta_{r} = \beta_{z} = 1.0 $,
the rotational speed increase as the toroidal magnetic field become larger.
The considerable matter about this panel is that, when  the toroidal magnetic field is very strong, the
rotational velocity will be Super-Keplerian (i.e., $ \beta_{\varphi}\sim 5 $). Moreover the surface density
of the disk decreases when the toroidal magnetic field become large. In figure
\ref{beta-phi}, the isothermal sound speed of the disk has distinctive properties between
$  \beta_{\varphi}<1 $ and $ \beta_{\varphi}>1 $. When $ \beta_{\varphi} $ is below
unity, the isothermal sound speed decrease, although for $ \beta_{\varphi} > 0 $, it increases.
It means that the is a minimum  near $ \beta_{\varphi} \sim 1$, because
the parameters $ \beta_{r} = \beta_{z} (=1.0) $ are present in our calculation and both of them decrease
the isothermal sound speed. So when the toroidal magnetic filed dominate, sound speed start be become large.

\begin{figure*}[tb]
\centering
\includegraphics[width=15cm]{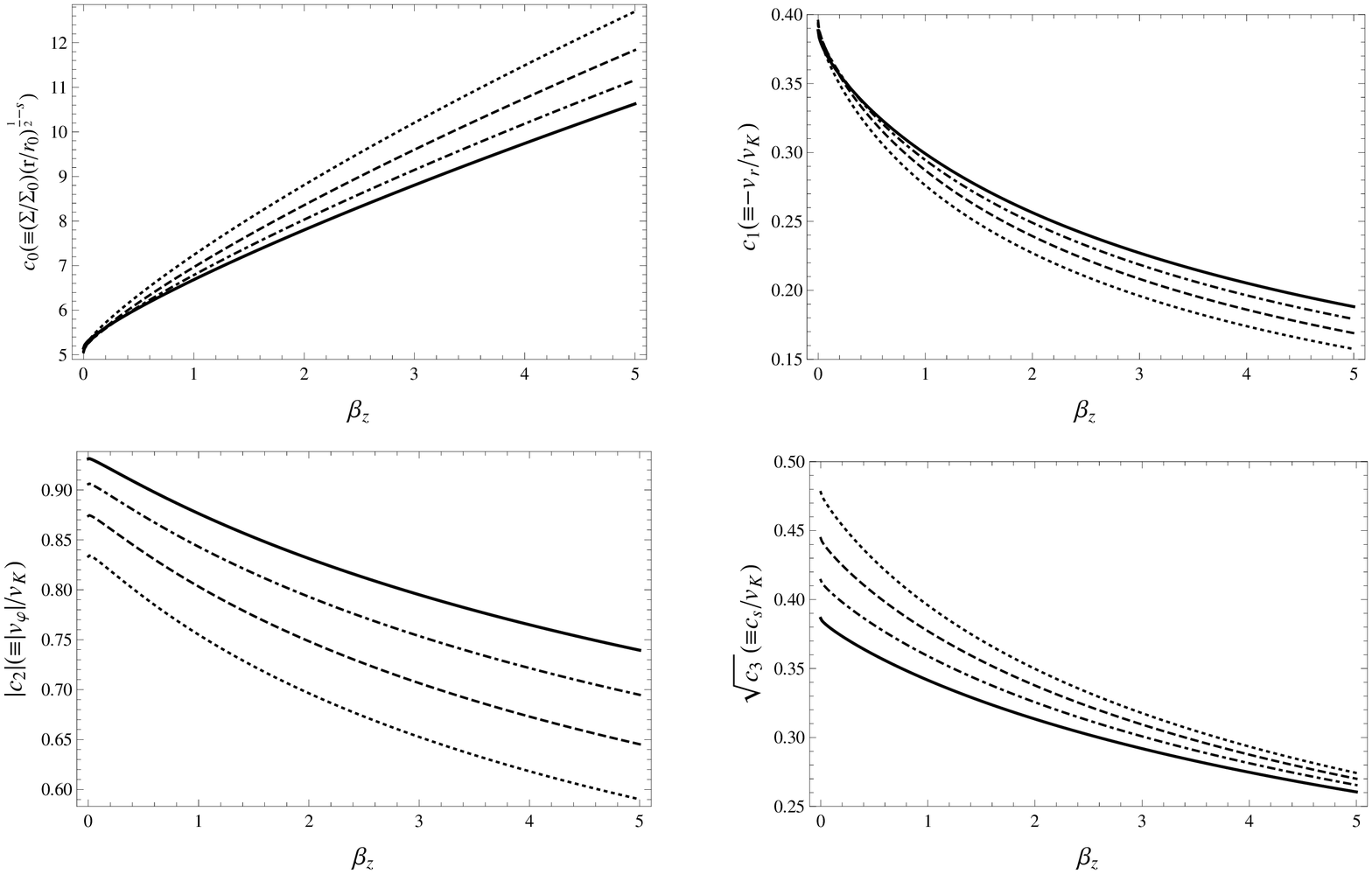}
\caption{Numerical coefficient $ c_{i} $s as a function of magnetic parameter
$ \beta_{z} $ for several values of $ s $. The dotted, dashed, dash-dotted and solid
lines correspond to $ s = 0.0, 0.1, 0.2 $ and $ 0.3 $ respectively. Parameters
are set as $ \alpha = 0.2 $, $ \gamma=4/3 $, $ \beta_{r} = \beta_{\varphi} = 1.0 $,
$ \phi_{s} = 0.01 $, $ \eta =  l = 0.1 $ and $ f = 1 $.}
\label{beta-z}
\end{figure*}

In figure \ref{beta-z} we have shown that the coefficients $ c_{i} $s with vertical component of magnetic field
for several values of $ s $. We figured out that a strong
$ (\beta_{z}) $ leads to an decrease of the accretion velocity $ v_{r} $, rotational speed $ v_{\varphi} $ as well as
isothermal sound speed $ c_{s} $ of the disc. That is because of  a large
magnetic pressure in the $ z $ direction prevents the accretion flow from being accreted, and will decreases
the effect of gas pressure as accretion proceeds. Although the surface density of the disk increases by
increasing of $ z $-component of magnetic field.

\section{General Properties of Accretion Flows }
In previous section it has introduced self-similar solutions and their properties for a
advection dominated accretion flow bathed in general large-scale magnetic field with thermal conduction and outflow/wind.
Having this self-similar solutions allow us to investigate the general properties of the disk, so we can see how
thermal conduction and wind/outflow will effect on the physical quantities of the disks.

The mass-accretion rate in the disk can be written by
a simple lapse function of radius due to the effect of outflow and thermal conduction. Using the self-similar solution we can have the mass accretion
rate as,
\begin{equation}
\dot{M}=-2\pi r \Sigma v_r= 2\pi c_0 c_1 \Sigma_0 (\frac{GM_*}{r_0})^{\frac{1}{2}}(\frac{r}{r_0})^s
=\dot{M}_{out}(\frac{r}{r_{out}})^s
\end{equation}
where $r_{out}$ and $\dot{M}_{out}$ are the outer's radius and accretion rate respectively. As we
have in last section thermal conduction, wind and large scale B-field will effect on $c_0, c_1$ through their parameters
$s, \phi$ and $\beta_{r, \phi, z}$. In the case of an accretion disk with no wind, $ s=0 $,
the accretion rate is independent of the radial distance, while for those with
wind/outflow, $ s > 0 $, the accretion rate decreases with radius as we
expect. Because winds start from different radii, the mass-loss rate
is not fixed and it depends on the radius. Then, some region of
accretion flow are not concentrated at the center, but are expand
over a wide space.

If the central star is a black hole of mass $M_*$, it will grows via accretion. It simply can estimate
the BH growth rate using accretion rate which also will affected by wind/outflow, thermal conduction and large-scale B-field, implicitly.

We can see the radial behavior of temperature of the present
self-similar advection dominated accretion discs with a general large-scale magnetic field, thermal conduction and outflow.
Optical depths of a hot accretion flow are highly dependent on their accretion rate. These ADAFs occur in two regimes which depend on optical depth and their mass accretion
rate. In a high mass accretion
rate, the optical depth becomes very high and the radiation
generated by accretion flow can be trapped into the disc. This
type of accretion disc is known as slim disc or optically thick disc.
In the low mass-accretion rates, the disc becomes optically
thin and the cooling time of accretion flow is longer than
the accretion time-scale. So, the generation energy by accretion flows
remains mostly in the discs and the discs are not able to radiate
their energy efficiently. Then, we should expect that the Large-scale magnetic field will effect on the
optical depth and accretion rate implicitly through the coefficient $c_0, c_1$.

We try to show that the radiative appearance of optically
thin discs, (such as $L_{\nu}$) , is not able to be calculated in ease. Also it should be
demonstrated the significance of advective cooling
in optically thin discs. To clarify the matter, we have to state 
how cooling and heating change with  $f$, $ r $ and $\dot{M}$. In all, the radiative
cooling was so complicate in case of dependency parameters. In
optical thin discs, the gas emission does not have a black body
continuum. Spectra emission seems to be ruled by Bremsstrahlung cooling by non relativistic or relativistic,
synchrotron and also Compton cooling. 
to deal will the optically thick case, where we are trying to study it, the radiation pressure dominates and
sound speed is related to radiation pressure. Then, we write the
average flux $ F $ as:
\begin{equation}
F=\sigma T_c^4=\frac{3c}{8H}\Pi_{gas}=\frac{3c}{8}\Sigma_0\frac{c_0c_3}{c_4}(\frac{GM_*}{r_0^2})(\frac{r}{r_0})^{s-\frac{5}{2}}
\end{equation}
Here, $\Pi_{gas}= \Sigma c_s^2$ is the height-integrated gas pressure, $T_c$ is the disc
central temperature and $\sigma$ is constant of the Stefan-Boltzmann .
in order to calculate the disk optical thickness in vertical direction of the optically thick case we can write:
\begin{equation}
\tau=\frac{1}{2}\kappa\Sigma=\frac{1}{2}\Sigma_0 c_0 (\frac{r}{r_0})^{s-\frac{1}{2}}
\end{equation}
where $\kappa$ is the electron-scattering opacity. Then, we calculate the
effective flux of the disc surface as:
\begin{equation}
\sigma T_{eff}^4=\frac{\sigma T_c^4}{\tau}=(\frac{3}{16\pi\sigma}\frac{c_3}{c_4}L_E)^{\frac{1}{4}}r^{-\frac{1}{2}}
\end{equation}
Here, $L_E $  is the Eddington luminosity of the central
object which is define as $L_E = 4\pi c\frac{GM}{\kappa}$. If we radially integrate these equations, we can have the disc
luminosity as,
\begin{equation}
L_{disc}=\int_{r_{in}}^{r_{out}}2\sigma T_{eff}^4 2\pi r dr=\frac{3}{4}L_E\frac{c_3}{c_4}\ln(\frac{r_{out}}{r_{in}})
\end{equation}
we can not emphasize more that the effective temperature
and the luminosity of the disc are not influenced by the magnetic field, thermal conduction and either mass-loss
through outflow (there are no $\beta_{r, \phi, z}$, $\phi_s  $ and $ s $ dependencies). But wind/outflow might affect
the radiative appearance of the disc through the $c_3$ and $c_4$ in these
formulae, implicitly. When we have mass-loss wind/outflow compared with the case of no
mass-loss, then the average flux decline through the disc
. For the mass-loss case, the surface density and therefore optical depth decrease
and we can see the deep inside of the disc.

\section{Conclusions}
\label{sect:conclusion}

In this paper, we studied an accretion disk in the advection dominated regime by considering a large-scale magnetic field and in presence of wind/outflow and  thermal conduction . Also, some approximations were made to simplify our main equations.
We supposed an static and axially symmetric disc with the $\alpha$-prescription of viscosity, $ \nu = \alpha c_{s} H $
and a set of similarity solution was presented for such a configuration.
We have extended Akizuki \& Fukue 2006; Zhang \& Dai 2008 and Abbassi et al 2010 self
similar solutions to present the structure of the advection dominated
accretion flows (ADAFs). Also, we ignored self-gravity of the discs and the general relativistic
effects.

Our results have shown that strong wind/outflow can have lower temperature
and there are satisfied with the results presented by Kawabata \&
Mineshige 2009. The most important
finding of these similar solutions is that, our accreting flow
is affected not only by mass-loss but also by loss of energy
by the wind/outflow.

There are some limitations in these solutions. One of them is that the accretion flow with conduction is a single-temperature structure. So, if one uses a two-temperatures structure for the ions and electrons in the discs, it is expected that the ions and electron temperatures decouple in the inner edge, which will modify the role of the conduction. The anisotropic character of  conduction in the presence of magnetic field is the other limitation in our solution (Balbus  2001)

Although our preliminary similarity solutions are so simplified, they obviously improve our understanding of the physics of hot accretion flow around a compact object. To have a more realistic picture of a hot accretion flow, a global solution is needed rather than the self similar solution. In our future studies we intend to investigate the effect of thermal conduction on the observational appearance and properties of a hot magnetized flow.

\label{lastpage}

\end{document}